\documentclass[conference]{IEEEtran}
\IEEEoverridecommandlockouts
\usepackage{cite}
\usepackage{amsmath,amssymb,amsfonts}
\usepackage{algorithmic}
\usepackage{graphicx}
\usepackage{textcomp}
\usepackage{xcolor}
\def\BibTeX{{\rm B\kern-.05em{\sc i\kern-.025em b}\kern-.08em
    T\kern-.1667em\lower.7ex\hbox{E}\kern-.125emX}}
    
\usepackage{color}

\usepackage{url}

\usepackage{booktabs}
\usepackage{colortbl} 
\usepackage{arydshln}
\usepackage{subfigure}
\usepackage{multirow}
\usepackage{multicol}
\usepackage{tablefootnote}
\usepackage{enumerate}
\usepackage{enumitem}
\usepackage[colorlinks, linkcolor=black, anchorcolor=black, citecolor=black, urlcolor=black]{hyperref}
\begin{document}

\title{Dual Graph Embedding for Object-Tag Link Prediction on the Knowledge Graph
}

\author{\IEEEauthorblockN{Chenyang Li\textsuperscript{1}, Xu Chen\textsuperscript{1}, Ya Zhang\textsuperscript{*1}, Siheng Chen\textsuperscript{2}, Dan Lv\textsuperscript{3}, and Yanfeng Wang\textsuperscript{1}}
\IEEEauthorblockA{\textsuperscript{1} \textit{Cooperative Medianet Innovation Center, Shanghai Jiao Tong University, Shanghai, China}\\
\textsuperscript{2} \textit{Mitsubishi Electric Research Laboratories, Cambridge, MA, USA}\\
\textsuperscript{3} \textit{StataCorp LLC, College Station, TX, USA}\\
\{lichenyanglh, xuchen2016, ya\_zhang\}@sjtu.edu.cn, sihengc@andrew.cmu.edu, dlv@stata.com, wangyanfeng@sjtu.edu.cn
}
}
\maketitle

\begin{abstract}
Knowledge graphs (KGs) composed of users, objects, and tags are widely used in web applications ranging from E-commerce, social media sites to news portals. This paper concentrates on an attractive application which aims to predict the object-tag links in the KG for better tag recommendation and object explanation. When predicting the object-tag links, both the first-order and high-order proximities between entities in the KG propagate essential similarity information for better prediction. Most existing methods focus on preserving the first-order proximity between entities in the KG. However, they cannot capture the high-order proximities in an explicit way, and the adopted margin-based criterion cannot measure the first-order proximity on the global structure accurately. In this paper, we propose a novel approach named Dual Graph Embedding (DGE) that models both the first-order and high-order proximities in the KG via an auto-encoding architecture to facilitate better object-tag relation inference.  Here the dual graphs contain an object graph and a tag graph that explicitly depict the high-order object-object and tag-tag proximities in the KG. The dual graph encoder in DGE then encodes these high-order proximities in the dual graphs into entity embeddings. The decoder formulates a skip-gram objective that maximizes the first-order proximity between observed object-tag pairs over the global proximity structure. With the supervision of the decoder, the embeddings derived by the encoder will be refined to capture both the first-order and high-order proximities in the KG for better link prediction. Extensive experiments on three real-world datasets demonstrate that DGE outperforms the state-of-the-art methods.
\end{abstract}


\section{Introduction} 
\label{sec:1}
In Web applications such as E-commerce and social media sites, many recommender systems incorporate the knowledge graph (KG) composed of users, objects, and tags to provide accurate and explainable recommendation~\cite{wang2019kgat}.
This paper focus on an attractive application which aims to predict the object-tag links in this kind of KG for better tag recommendation~\cite{Belem:2017} and object management~\cite{Tso-Sutter:2008}. 

\begin{figure}[tp]
    \centering
    \includegraphics[width=1.\linewidth]{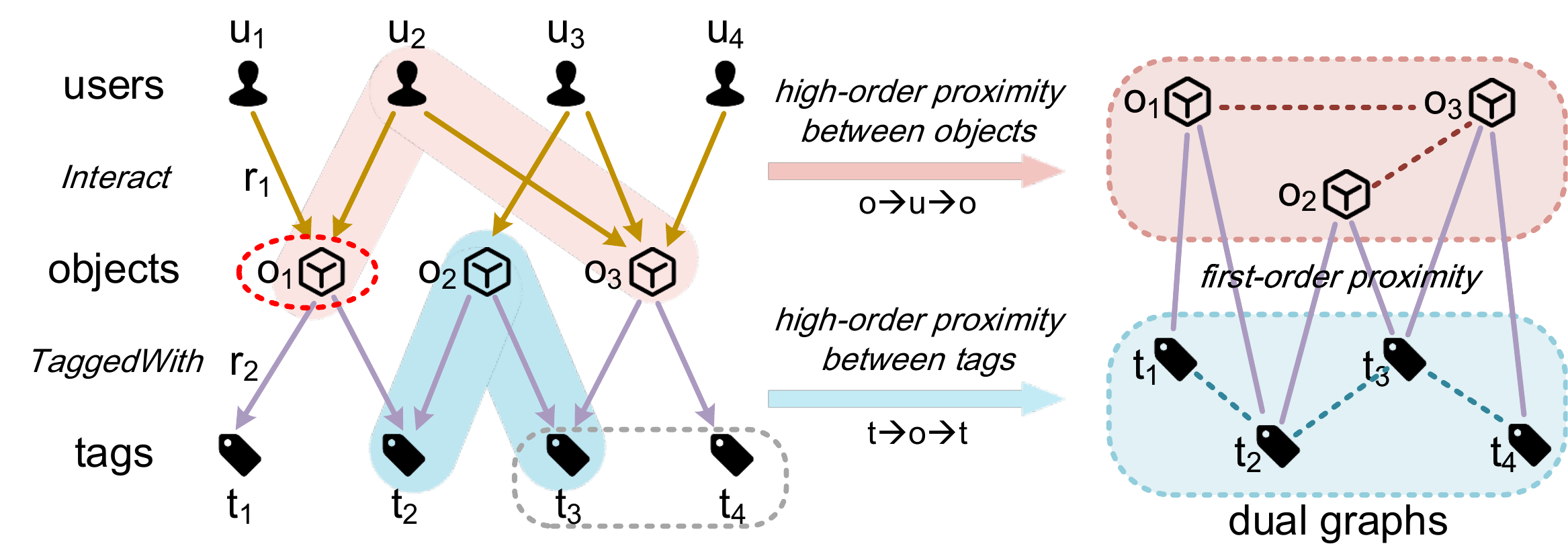}
    \caption{A toy example of the KG in the object-tag link prediction task. The purple lines indicate the first-order object-tag proximity. 
    The high-order proximity exists between the objects in the left red region. Tags in the left blue region also share the high-order proximity. These high-order proximities are depicted in the dual graphs shown on the right side.
    $o_1$ with the red circle is the head whose tails need to be predicted.
    Tags in the grey circle are discovered by high-order relationships in the KG.}
    \label{fig:intro}
    \vspace{-6pt}
\end{figure}

In the object-tag link prediction problem, the knowledge graph contains three types of entities (i.e., users, objects, and tags) and two types of relations (i.e., Interact and TaggedWith) as Fig.~\ref{fig:intro} shows. Two types of (head, relation, tail) 
triplets with the forms of (user, Interact, object) and (object, TaggedWith, tag) are included in the KG. In particular, this task focuses on the latter type of triplets by taking the objects as the heads and the tags as the tails to predict. 
Here both the first-order and high-order proximities in the KG provide structural similarity information to enhance the link prediction. 
At first, the first-order proximity between a head-tail pair determines the existence of the corresponding link directly.
In Fig.~\ref{fig:intro}, the object and the tag that are associated by an observed link of type $r_2$ share the stronger first-order proximity than the nodes that are not directly linked. The existing first-order proximity information in these observed links can discover
new links between objects and tags that share high proximity.
Besides the first-order proximity, the high-order object-object and tag-tag proximities implied in the high-order connectivities in the KG provide collaborative signals for link prediction.
More specifically, the high-order proximity between two objects encourages one of the objects to link to the tags of the other one. 
Similarly, the high-order proximity between two tags enriches the links for the object linked to any one of these tags.
For example, in Fig.~\ref{fig:intro}, $o_1$ and $o_3$ are two-hop neighbors on the path $o_1\xrightarrow{-r_1}u_2\xrightarrow{r_1}o_3$, and they share high-order proximity. Then $t_4$ that is linked to $o_3$ may be also relevant to $o_1$.
Similarly, the tag $t_2$ of $o_1$ shares the high-order proximity with $t_3$ over the path $t_2\xrightarrow{-r_2}o_2\xrightarrow{r_2}t_3$, and a link between $o_1$ and $t_3$ is probable to be added. 
In this sense, both the first-order and high-order proximities in the KG need to be captured for high-quality link prediction.

However, existing models fail to capture both the first-order and high-order proximities in the KG jointly.
Translational distance models designed for KG completion measure the distances between entities and their immediate neighbors after the translations carried by the relations~\cite{bordes2013transe,Wang2014transh,Lin2015transr}.
In this way, these models concentrate on the first-order proximity between existing triplets~\cite{Ji2020ASO}.
Some other methods based on random walk~\cite{Jiang:2018} or feature aggregation~\cite{vdberg2017graph,Wang:2019} on the KG focus on the information propagation from current entity to its directly linked entity.
Without capturing the high-order proximities explicitly, these methods will lose the collaborative information in the high-order connectivities in the KG, which is essential for prediction with high accuracy. For example, $t_3$ and $t_4$ in Fig.~\ref{fig:intro} may be overlooked by these methods when predicting the tails of $o_1$.
Besides, the margin-based criterion commonly used in the KG completion methods only considers the pairwise relations between heads and tails, which may fail to depict the first-order proximity between any pair of entities over the global proximity structure.

In this paper, we propose a Dual Graph Embedding (DGE) model in an auto-encoding architecture that simultaneously captures the first-order and high-order proximities in the KG to predict the missing object-tag links.
Here the dual graphs contain an object graph and a tag graph constructed based on the high-order connectitities in the KG. Hence links in the dual graphs describe the high-order object-object and tag-tag proximities explicitly.
The dual graph encoder in DGE then captures these high-order proximities by encoding the structural information in the dual graphs into entity embeddings.
In the decoder, instead of the widely-used margin loss,
we formulate a skip-gram objective that maximizes the likelihood that each observed tag is relevant to a given object over all the possible tags.
In this way, the decoder measures
the first-order proximity in the global proximity structure in the KG and refines the embeddings learned from the encoder. The auto-encoding formulation encourages DGE to capture the first-order and high-order proximities in an end-to-end manner for better prediction.
We conduct our experiments on several real-world datasets for tag recommendation tasks. The results show that DGE
predicts high-relevant object-tag pairs compared to 
the state-of-the-art methods. 
Our contributions could be summarized as follows:
\begin{itemize}
    \item We propose a Dual Graph Embedding (DGE) method to capture both the first-order and high-order proximities in the KG simultaneously, and further improve the quality of the object-tag link prediction;
    \item We adopt the skip-gram objective that maximizes the likelihood of the observed tags given an object over the candidate tag set, and the first-order proximity between each object-tag pair in the global proximity structure can be measured more accurately;
    \item Extensive experiments are conducted on three real-world datasets for tag recommendation. The empirical results show that our method outperforms the state-of-the-art methods on relevant tag prediction for target objects.
\end{itemize}

\section{Related Work} \label{sec:2}
\subsection{Knowledge Graph Completion} 
Considering the incompleteness of knowledge graphs, many methods have been proposed to add new triplets to the knowledge graph. A typical task of KG completion is link prediction, which is widely applied to recommender systems in real-life scenarios~\cite{Ji2020ASO}.

Many translational distance models learn low-dimension embeddings of entities and relations by minimizing the distance between two directly linked entities in a translated space~\cite{Wang2017}. Besides, they adopt the margin-based criterion to measure the first-order proximity in the KG.
TransE~\cite{bordes2013transe} treats relations as translations from heads to tails. This method supposes that the added result of the head and relation embeddings should be close to the tail embedding. To follow up, TransH~\cite{Wang2014transh} modifies the scoring function by projecting entities and relations into a hyperplane. TransR~\cite{Lin2015transr} introduces relation-specific spaces and projects the head and tail embeddings into the corresponding space. To simplify TransR, TransD~\cite{Ji2015transd} replaces the projection matrix into the product of two mapping vectors. Besides, some methods like KG2E~\cite{He2015KG2E} and TransG~\cite{xiao2016transg} redefine the distance by assuming that the entity and relation embeddings are from Gaussian distributions.


\subsection{Tag Recommendation} 
Tag recommendation methods can be formulated as the link prediction problem on the user-object-tag relation graph~\cite{Belem:2017}.

Some matrix factorization (MF) based methods model the pairwise relationships among users, objects and tags to learn the low-dimension embeddings~\cite{Rendle:2010}.
Other methods additionally extract the tag co-occurrences and model the tag-tag proximity to recommend relevant tags for certain objects~\cite{Rae:2010,Liang:2016}.
Some recent methods employ the random walk with restart or node feature aggregation scheme in the input graph to predict the link between object nodes and tag nodes~\cite{Jiang:2018,vdberg2017graph,Wang:2019}. Although they consider the high-order connectivities in the input graph, they focus on the information propagation process between two directly linked nodes.

In summary, methods for KG completion and tag recommendation cannot capture the collaborative signals from the high-order object-object and tag-tag proximities in the KG explicitly. Besides, the margin loss employed for KG completion cannot measure the first-order proximity over the global proximity structure accurately.

\begin{figure*}[htbp]
\includegraphics[width=.9\linewidth]{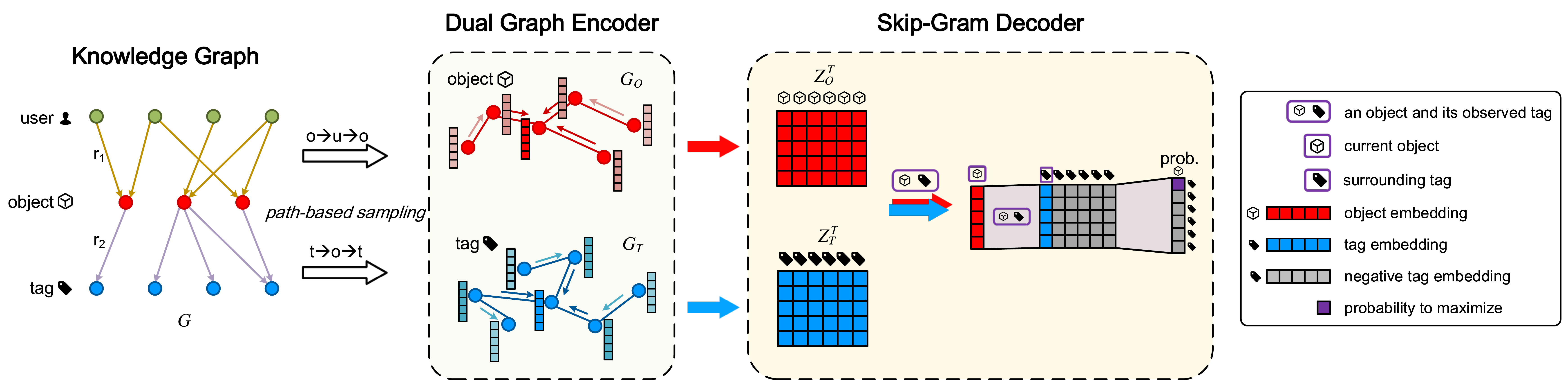}
\centering
\caption{The Dual Graph Embedding model for the prediction of the object-tag links. We adopt the path-based sampling scheme 
to construct the object and tag graphs that contain the high-order proximities. A dual graph encoder contains two 2-layer GCN architectures to embed the object-object and tag-tag proximities. The decoder measures the first-order proximity between objects and tags, and uses the skip-gram objective to supervise the encoder.
}
\label{f1}
\vspace{-4pt}
\end{figure*}

\section{Preliminary}
\subsection{Problem Definition}
In the object-tag link prediction task, we first introduce the input knowledge graph and the task goal.

\textbf{Knowledge graph in this task.}
We denote the knowledge graph in this task as $G=\{(h,r,t)|h,t\in\mathcal{E},r\in\mathcal{R}\}$, where $\mathcal{E}=\mathcal{U}\cup\mathcal{O}\cup\mathcal{T}$. Here $\mathcal{U}=\{u_1,u_2,\cdots,u_P\}$, $\mathcal{O}=\{o_1,o_2,\cdots,o_N\}$, and $\mathcal{T}=\{t_1,t_2,\cdots,t_M\}$ are the user, object and tag set with $P$, $M$, and $N$ entities respectively. The relation set $\mathcal{R}$ is composed of two types of relations including $r_1=Interact$ and $r_2=TaggedWith$ as Fig.~\ref{fig:intro} shows.

\textbf{Task description.}
Given the input $G$, we aim to predict the existence of the object-tag link between any pair in $\{(h,t)|h\in\mathcal{O},t\in\mathcal{T}\}$. Here objects are considered as given heads and tags are treated as candidate tails to predict.
In particular, we denote the observed tag set of a given object $o_i$ as $\mathcal{S}_{o_i}=\{t^i_1,t^i_2,...,t^i_{M_i}\}$, where $\mathcal{S}_{o_i}$ contains $M_i$ observed tags linked to $o_i$ and $\mathcal{S}_{o_i}\subset \mathcal{T}$. 
\subsection{Terminology Explanation}
We give formal definitions of some important terminologies
related to this task:

\textbf{High-order connectivity.}
We take the definition given in~\cite{wang2019kgat} that the $L$-order connectivity means the long range path $e_0\xrightarrow{r^1}e_1\xrightarrow{r^2}\cdots\xrightarrow{r^L}e_{L}$, where $e_i\in\mathcal{E}$ and $r^j\in\mathcal{R}$ for $i=0,\cdots,L$ and $j=1,\cdots,L$. Then $e_{L}$ is the $L$-hop neighbor of $e_0$.

\textbf{Path-based neighbors.}
According to the high-order connectivities, we consider a predefined path set
$\mathcal{P}=\{e_0\xrightarrow{r^1}\cdots\xrightarrow{r^L}e_{L}\}$ by fixing the types of all the entities and relations as well as forcing $e_0$ and $e_{L}$ to be the same types. Then for any path $pa\in\mathcal{P}$,
the start and end entities are $\mathcal{P}$-path-based neighbors of each other.

\textbf{High-order object-object and tag-tag proximities.}
The high-order proximity exists between an entity and its path-based neighbor given a predefined $\mathcal{P}$. 
We focus on the proximities existing in objects and tags since this task only predicts the missing object-tag links. 
Given an object $o_i$, $o_i$ and $o_j$ share high-order proximity if $o_j$ is a path-based neighbor of $o_i$. The high-order tag-tag proximity can be defined in the same way.

\section{Proposed Model}
\label{sec:4}

\subsection{Model Overview}
Our proposed DGE consists of a dual graph encoder and a skip-gram decoder, and the model framework is depicted in Fig.~\ref{f1}. 
The dual graphs here indicate an object graph and a tag graph containing the high-order object-object and tag-tag proximities in the KG respectively.
The encoder then extracts both the high-order proximities from the dual graphs and embeds them into entity embeddings.
Besides, the skip-gram decoder measures the first-order proximity over the global proximity structure in the KG
to determine the existence of links and to supervise the entire model.

According to the high-order connectivities in the input KG, we can sample the path-based neighbors of objects and tags respectively. These extracted neighboring correlations that contain the high-order proximities are utilized to build the object graph $G_O=(V_O,E_O)$ and the tag graph $G_T=(V_T,E_T)$, where $V_T$ and $E_T$ denote the vertex set and link set of tags (similarly, $V_O$ and $E_O$ are for objects). Both graphs form the dual graphs, where we assume that the directional information in the initial KG is not important in this task.
Then the encoder extracts the high-order object-object and tag-tag proximities from the dual graphs $G_O$ and $G_T$ respectively. The encoded object and tag embeddings are given by
\begin{equation}
    \label{e_enc}
    Z_O, Z_T=DGEnc(G_O, G_T),
\end{equation}
where $Z_O\in \mathbb{R}^{N\times d}$ and $Z_T\in \mathbb{R}^{M\times d}$ are the object and tag embeddings respectively with the latent dimension $d$, and $DGEnc$ denotes the encoding process. 

In the decoder, for an object $o_i$, we assume that the observed tags in $\mathcal{S}_{o_i}$ share the stronger first-order proximity
with $o_i$ compared to those globally unobserved ones. Actually, this assumption is consistent with the idea of the skip-gram model~\cite{mikolov2013distributed} that the \emph{surrounding} words are more related to the \emph{current} word compared to those globally distant words.
In this sense, to capture the dominant proximity information in the global structure
for any $o_i$, we formulate the decoder from the skip-gram perspective.
More specifically, given any pair $(o_i, t_j)$, the \emph{current} object vector is the $i$th row of $Z_O$ denoted by $Z_{O,i}$ and the target tag vector is the $j$th row of $Z_T$ denoted by $Z_{T,j}$.
Here $Z_T$ is taken as the candidate \emph{surrounding} tag embedding matrix.
Then the decoder calculates the probability $p(t_j|o_i)$ that $t_j$ is linked to $o_i$ in the same way as the skip-gram model based on the indexed and candidate vectors,
\begin{equation}
    \label{e_dec}
    p(t_j|o_i)=SGDec(Z_{O,i},Z_{T,j}),
\end{equation}
where $SGDec$ implies the skip-gram decoding process. 
To supervise the whole model, for the \emph{current} $o_i$, the skip-gram objective~\cite{mikolov2013distributed} maximizes the likelihood of its $M_i$ observed \emph{surrounding} tags in $\mathcal{S}_{o_i}$, which is given by
\begin{align}
    \label{e12}
 &\max p(t^{i}_1,t^{i}_2,\cdots,t^{i}_{M_i}|o_i) \nonumber\\
 =&\max\prod_{m=1}^{M_i}p(t^{i}_m|o_i), \\
    =&\max\prod_{m=1}^{M_i} SGDec(DGEnc(G_O,G_T)[o_i,t^i_m]).  \nonumber
\end{align}
During the training process, $Z_O$ and $Z_T$ will be refined to capture more similarity information from both the first-order and high-order proximities in the KG to assist the prediction.

\subsection{Dual Graph Encoder}
Since the high-order proximities in the input knowledge graph $G$ propagate essential information for link prediction, we introduce a dual graph encoder to capture these proximities.
More specifically, we construct the dual graphs including the object graph $G_O=(V_O,E_O)$ and the tag graph $G_T=(V_T,E_T)$ via a path-based neighbor sampling scheme on the input graph $G$. Thus links in both graphs illustrate the high-order proximities in the input $G$.
Then the dual graph encoder embeds the structural information of $G_O$ and $G_T$ into the embeddings $Z_O$ and $Z_T$. In this way, $Z_O$ and $Z_T$ fetch the two types of high-order proximities to provide collaborative signals for the subsequent decoder to predict the object-tag links.
The internal layout of the encoder is given in Fig.~\ref{f1}.

\textbf{Encode the high-order proximity between objects.}
In this process, we encode the object graph $G_O$ to mine the high-order proximity between objects. 
We consider the path set $\mathcal{P}_o=\{o_i\xrightarrow{-r_1}u_k\xrightarrow{r_1}o_j|o_i,o_j\in\mathcal{O},u_k\in\mathcal{U}\}$ in the input $G$, and the $\mathcal{P}_o$-path-based neighbors can be sampled for each object node. Note that the path set $\mathcal{P}_o'=\{o_i\xrightarrow{r_2}t_k\xrightarrow{-r_2}o_j|o_i,o_j\in\mathcal{O},t_k\in\mathcal{T}\}$ is not selected since the information propagated on $\mathcal{P}_o'$ can be captured when representing the tag graph. Accordingly, $G_O$ is constructed by adding the link between each pair of the sampled object nodes. To depict the semantic similarities between any two nodes,
we utilize Sparse Positive PMI (SPPMI)~\cite{levy2014neural} values which are commonly used in NLP tasks to normalize the link weights of $G_O$. Then we denote the corresponding adjacency matrix as $A_O\in \mathbb{R}^{N\times N}$.

Considering the superior performance of Graph Convolutional Networks (GCNs)~\cite{kipf2017semi} on capturing relations between nodes, we apply a two-layer GCN to encode the object graph's information into $Z_O$. When the content features $X_O\in \mathbb{R}^{N\times d_{in}}$ with the feature dimension of $d_{in}$ are provided, GCN can extract the information from both the graph topological structure and the input features. Otherwise, the content features are one-hot encodings in $N$ dimension and $X_O$ equals to the $N$-by-$N$ unit matrix $I_N$. GCN still represents the structural information at this time. Since the datasets contain no specific content features, we adopt one-hot encodings as the node features in our experiments. Thus, the two-layer GCN encoder for $G_O$ is given by
\begin{equation}
    \label{e:gcn_o}
    Z_O= \hat{A}_O\mathrm{ReLU}(\hat{A}_OX_OW_O^{(0)})W_O^{(1)},
\end{equation}
where the normalized 
$\hat{A}_O=\bar{D}_O^{-\frac{1}{2}}\bar{A}_O\bar{D}_O^{-\frac{1}{2}}$ with $\bar{A}_O=A_O+I_N$ and $\bar{D}_{O(ii)}=\sum_j \bar{A}_{O(ij)}$. Besides, $W_O^{(0)}\in\mathbb{R}^{d_{in}\times h}$ and $W_O^{(1)}\in\mathbb{R}^{h\times d}$ are weight matrices for the first and second layers of GCN respectively. $Z_O\in \mathbb{R}^{N\times d}$ is the output object embedding matrix representing the local graph structure and the node features if provided. Being split by rows, 
$Z_O=\{Z_{O,i}, i=1,\cdots,N\}$ is then fed to the decoder for prediction.

\textbf{Encode the high-order proximity between tags.} 
Similarly, we encode the tag graph $G_T$ to extract the high-order proximity between tags. $G_T$ is constructed based on the $\mathcal{P}_t$-path-based neighbor sampling process, where $\mathcal{P}_t$ in the input $G$ is defined as $\{t_i\xrightarrow{-r_2}o_k\xrightarrow{r_2}t_j|t_i,t_j\in\mathcal{T},o_k\in\mathcal{O}\}$. The link weight between each pair of tag nodes in $G_T$ is determined by the SPPMI value of them, and the adjacency matrix $A_T \in \mathbb{R}^{M\times M}$ is composed of all the SPPMI values. Then the high-order proximity between tags are encoded into tag embeddings via another two-layer GCN encoder,
\begin{equation}
    \label{e:gcn_t}
    Z_T= \hat{A}_T\mathrm{ReLU}(\hat{A}_TW_T^{(0)})W_T^{(1)},
\end{equation}
where $\hat{A}_T$ is normalized in the same way as $\hat{A}_O$, and $W_T^{(0)}\in\mathbb{R}^{M\times h}$, $W_T^{(1)}\in\mathbb{R}^{h\times d}$ are weight matrices of GCN layers. Since no certain features for tags are provided,
we omit the input features $X_T$ here. $Z_T\in \mathbb{R}^{M\times d}$ is the output tag representation capturing the proximity structure of tag graph. $Z_T=\{Z_{T,j}, j=1,\cdots,M\}$ are considered as all candidate \textit{surrounding} tag embeddings for any objects in the decoder.

\subsection{Skip-Gram Decoder}
In the skip-gram decoder, we measure the first-order proximity between objects and tags over the global proximity structure to predict the missing object-tag links.
Besides, for each object, the decoder maximizes the likelihood of the corresponding observed tags to supervise the entire model. The decoding process is shown in Fig.~\ref{f1}.

Given the object embeddings $Z_O$ and the tag embeddings $Z_T$, 
the first-order proximity between
the \emph{current} $o_i$ and the target $t_j$ is measured by the inner product of their corresponding embeddings, i.e., the relevance score is
\begin{equation}
\label{e10}
    s(o_i,t_j)={(Z_{T,j})}^T Z_{O,i},
\end{equation}
where $Z_{O,i}$ and $Z_{T,j}$ are indexed from $Z_O$ and $Z_T$ respectively.
The probability $p(t_j|o_i)$ for $t_j$ and $o_i$ is then calculated by applying the softmax operation over the candidate set~\cite{mikolov2013distributed},
\begin{equation}
    \label{e11}
    p(t_j|o_i)=\mathrm{softmax}\left(s(o_i,t_j)\right)=\frac{\exp s(o_i, t_j)}{U(o_i)},
\end{equation}
where $U(o_i)=\sum_{k=1}^M \exp ({(Z_{T,k})}^T Z_{O,i})$ represents the normalization factor of $o_i$. Here $U(o_i)$ contains the global proximity information for the head $o_i$.

During the optimization process, the probabilities in~\eqref{e12} can be calculated by~\eqref{e11}. By maximizing the likelihood for all the cases of each object with its observed tag set, the object and tag embeddings will be refined to get more similarity information
from the first-order and high-order proximities to improve the prediction accuracy. 

\textbf{Sub-sampling.}
In the training process, 
we calculate the probability in~\eqref{e11} for every observed object-tag pair.
Unfortunately calculating~\eqref{e11} 
requires normalizing over the entire $\mathcal{T}$, which means that it is prohibitively expensive to train the model.
Inspired by a sub-sampling scheme of noise-contrastive estimation (NCE) which is widely used in word embedding models~\cite{Mnih2013}, we employ NCE to achieve fast training.
This scheme trains a binary classifier with label $y$ treating observed tags from data distribution 
$P_d^{o_i}$ 
as positive samples ($y=1$) and tags from a noise distribution $P_n$ as negative ones ($y=0$) given $o_i$. 
Assuming that the negative samples appear $K$ times more frequently than the positive ones, the probability that a given tag $t_j$ being a positive tag for 
$o_i$ is
\begin{align}
\label{e14}
 p(y=1|t_j, o_i)&=\frac{p(t_j|o_i)}{p(t_j|o_i)+K p_n(t_j)}\\
        \xrightarrow{\mathrm{omit}\, U(o_i)}&=\frac{\exp s(o_i,t_j)}{\exp s(o_i,t_j)+K p_n(t_j)},\nonumber
\end{align}
where $p_n(t_j)$ means $t_j$ from noise distribution, and 
the unnormalized model with a scaled noise distribution by ignoring $U(o_i)$ in~\eqref{e11} can be normalized 
during training~\cite{Mnih2013}.
Hence, 
the probability of $t_j$ being a negative sample for $o_i$ is 
$p(y=0|t_j, o_i)=1-p(y=1|t_j, o_i)$. 
Being consistent with the objective in~\eqref{e12}, the goal 
is converted to maximize the 
likelihood for the correct labels $y$, averaged over the positive and negative data sets. For 
$o_i$, the log-likelihood is
\begin{align}
         \mathcal{J}_\Theta(o_i)=& E_{P_d^{o_i}}[\log p(y=1|t_j, o_i)]\nonumber\\
          \label{e15}
         +&KE_{P_n}[\log p(y=0|t_j, o_i)] \\
         \approx&\log p(y=1|t_j, o_i)+\sum_{k=1}^K\log p(y=0|t_k, o_i).\nonumber
\end{align}
Here the expectations over the data and noise distributions are approximated by sampling during training~\cite{Mnih2013}.
Then the overall objective is the summation of the likelihood for all the objects, and the optimization goal can be presented as
\begin{equation}
    \label{e16}
        \max_{\Theta} \sum_{o_i\in \mathcal{O}} \mathcal{J}_\Theta(o_i),
\end{equation}
where $\mathcal{J}_\Theta(o_i)$ is obtained by~\eqref{e15}. Here the optimization parameters $\Theta=\{W_O^{(0)},W_O^{(1)},W_T^{(0)},W_T^{(1)}\}$ determine the encoder, namely the embedding process of the high-order object-object and tag-tag proximities.

\begin{table}[t]
    \centering
    \caption{summary Statistics of Three Datasets}
    \label{t2}
    \resizebox*{0.86\linewidth}{!}{ 
    \begin{tabular}{|c|ccc|}
    \hline
         & Movielens-1M & LastFm & Steam \\
        \hline\hline
        \#objects & 3,883 & 17,632 & 9,373 \\
        \#tags & 1,008 & 11,946 & 352 \\
        \#users & 6,040 & 1,892 & 101,654
        \\
        \hline\hline
        $\#r_1$ (Interact) & 1,000,209 & 70,297 & 1,100,628 \\
        density of u-o interactions & 4.2647\% & 0.2108\% & 0.1155\% \\
        sparsity of object graph & 4.5047\% & 0.8492\% & 0.3505\% \\
        \hline\hline
        $\#r_2$ (TaggedWith) & 15,498 & 108,437 & 83,700 \\
        density of o-t observations & 0.3969\% & 0.0515\% & 2.5369\% \\
        sparsity of tag graph & 3.2675\% & 0.7061\% & 3.8820\% \\
        \hline
    \end{tabular}
    }
\end{table}

\begin{table}
    \centering
    \caption{Experimental Settings ($k_O$ and $k_T$ control the sparsity of SPPMI matrices of object and tag graph respectively.)}
    \label{t:set}
    \resizebox*{.95\linewidth}{!}{
    \begin{tabular}{|c|l|}
    \hline
    Methods & Settings \\
    \hline\hline
    MF & hidden size $d=100$ \\
    \hline
    TransE & hidden size $d=100$ \\
    \hline
    TransH & hidden size $d=100$ \\
    \hline
    TransR & hidden size of entities $d_e=100$, hidden size of relations $d_r=100$\\
    \hline
    Skip-Gram & hidden size $d=100$ \\
    \hline \hline
     CoFactor & Movielen, Steam: $k$ for SPPMI$=1$, $d=16$;\\
     &LastFM: $k=2$, $d=64$\\
     \hline
    MAD & $\mu_1=\mu_3=1$, $\mu_2=1e-4$ \\
    \hline
    HeteLearn & $\alpha=0.8$ \\
    \hline
    GCMC+, & Movielen, Steam network structure: $\{128, 128, 128\}$;\\
    NGCF+ & LastFM neitwork structure: $\{64, 64, 64, 64\}$;\\
    \hline
    & batch size=64, epochs=300, learning rate=2e-3, \\
    &15 negative samples for each positive one,\\
     DGE & Movielen: $k_O=0.1, k_T=1$, 2-layer GCN structure: $\{32,16\}$;\\
     & LastFM: $k_O=1, k_T=1$, 2-layer GCN structure: $\{64,64\}$;\\
     & Steam: $k_O=5, k_T=5$, 2-layer GCN structure: $\{32,16\}$;\\
    \hline
    \end{tabular}
    }
\end{table}

\section{Experiments} \label{sec:5}
\subsection{Datasets}
We adopt three real-world tag recommendation datasets including Movielens-1M\footnote{\url{https://movielens.org/}}, LastFm\footnote{\url{https://www.last.fm/}} and Steam\footnote{\url{https://store.steampowered.com/}} to evaluate our model. We summarize the statistical information in Table~\ref{t2}.

\begin{itemize}
    \item Movielens-1M: 3,883 movies are rated by 6,040 users.
    Tags of each movie are chosen from all 1,008 tags. A total of 15,498 object-tag interactions are observed.
    
    \item LastFm: 17,632 artists are listened and tagged by 1,892 users, and the tags are from a tag set with a size of 11,946. In total, 108,437 object-tag observations are included.
    
    \item Steam: 9,373 apps are reviewed by 101,654 users. The size of the tag set is 352 and 83,700 object-tag pairs are observed.
\end{itemize}

\subsection{Baseline Methods.}
To demonstrate the effectiveness of the proposed DGE, we compare DGE with five methods only modeling the first-order proximity and five methods additionally modeling part of the high-order proximities.

\subsubsection{Methods based on the first-order proximity}
The five baseline models developed for the object-tag link prediction includes
\textbf{MF}~\cite{koren2009matrix}, \textbf{Skip-Gram}~\cite{Mnih2013}, \textbf{TransE}~\cite{bordes2013transe}, \textbf{TransH}~\cite{Wang2014transh}, and \textbf{TransR}~\cite{Lin2015transr}. MF factorizes the object-tag interaction matrix into two low-dimension feature matrix for objects and tags. It employs the mean square error (MSE) as the loss function. The skip-gram model adopts the same way as MF to generate features, but it utilizes NCE loss to measure the first-order proximity.
The latter three methods treat the relation embeddings as the translation embeddings and define the distance via different scoring functions. They use the margin loss in the training process. In this task, to avoid the bias to irrelevant links in the KG, we adopt the same way as \cite{Cao2019} to add an extra Bayesian personalized ranking loss to narrow the translation distance between the related objects and tags.


\subsubsection{Methods utilizing part of the high-order proximities}
We use five methods that model part of the high-order proximities as the comparison methods involving
\textbf{CoFactor}~\cite{Liang:2016}, \textbf{MAD}~\cite{Talukdar2009NewRA}, \textbf{HeteLearn}~\cite{Jiang:2018}, \textbf{GCMC}~\cite{vdberg2017graph} and \textbf{NGCF}~\cite{Wang:2019}. 
CoFactor based on MF and tag co-occurrences only considers the high-order proximity between tags in the KG.
MAD conducts label propagation on the object graph that reflects the object-object proximity. HeteLearn is a state-of-the-art method which provides tags via the random walk scheme on the user-object-tag graph. GCMC and NGCF predict links in the bipartite graph based on node feature aggregation. Since they do not distinguish node types during feature aggregation, we extend them to the user-object-tag tripartite graph denoted by \textbf{GCMC+} and \textbf{NGCF+}. 


\begin{table}[t]
    \centering
    \caption{$Recall@k$, $NDCG@k$ on Movielens-1M, LastFM, and Steam based on Different Methods}
    \label{tab:res}
    \resizebox*{1.\linewidth}{!}{
    \begin{tabular}{|c|c|cccc|}
        \hline
        Datasets& Methods & Recall@3 & NDCG@3 & Recall@5 &	NDCG@5 \\
        \hline\hline
         & MF & 0.7252 & 0.4116 &  0.7819 & 0.3088\\
         & TransE &	0.7292 &	0.4125 & 0.7776 &	0.3085\\
        & TransH  &	0.7293 &	0.4127 & 0.7760 &	0.3081\\
        & TransR &	0.6496 &	0.3390 & 0.7227 &	0.2597\\
        & Skip-Gram & 0.7467 & 0.4209 & 0.7895 & 0.3144 \\
        \cline{2-6}
        Movielens-1M & CoFactor &	0.7234 & 0.4038 &	0.7825 &	0.3043  \\
        & MAD &	0.7359 &	0.4144 &	0.7774 &	0.3097 \\
        & HeteLearn &	0.7863 &	0.4356 &	0.8290 &	0.3249 \\
        & GCMC+ &  0.8293&	0.4664 & 	0.8528&	0.3434 \\
        & NGCF+ & 0.7549 &	0.4242 & 0.7944 &	0.3162 \\
        \cline{2-6}
        & \textbf{DGE(ours)}  &	\textbf{0.8464} &	\textbf{0.4850} & \textbf{0.8677} & \textbf{0.3565}\\
        \hline\hline
        & MF & 0.0667 & 0.0601 & 0.0904 & 0.0536\\
        & TransE & 0.0746 & 0.0668 & 0.1100 & 0.0618  \\
        & TransH & 0.0751 & 0.0669 & 0.1101 & 0.0617 \\
        & TransR & 0.0637 & 0.0561 & 0.0819 & 0.0480 \\
        & Skip-Gram & 0.1322 & 0.1054 & 0.1920 & 0.0966\\
        \cline{2-6}
        LastFM & CoFactor &	0.1538 &	0.1277  &	0.1969 &	0.1073 \\
        & MAD &	0.2245 &	0.1927 &	0.2877 &	0.1605 \\
        & HeteLearn &	0.2376 &	0.1898 &	0.3119 &	0.1608 \\
        & GCMC+ & 0.1703 & 0.1310 & 0.2310 & 0.1151\\
        & NGCF+ & 0.1011 & 0.0775 & 0.1414 & 0.0703 \\
        \cline{2-6}
        & \textbf{DGE(ours)} &	\textbf{0.2494} &	\textbf{0.2129} &	\textbf{0.3154} &	\textbf{0.1772} \\
        \hline\hline
        & MF & 0.2756 & 0.2151 & 0.3642 & 0.1802\\
        & TransE &	0.2932 &	0.2225 & 0.3744 &	0.1859\\
        & TransH & 0.2915 &	0.2217 & 0.3748 &	0.1850 \\
        & TransR & 0.2805 &	0.2042 & 0.3678 &	0.1737\\
        &Skip-Gram & 0.3203 & 0.2386 & 0.4082 & 0.1999 \\
        \cline{2-6}
        Steam & CoFactor &	0.1538 &	0.1277  &	0.1969 &	0.1073 \\
        & MAD  &	0.3297 &	0.2539  &	0.4268 &	0.2133\\
        & HeteLearn  &	0.3856 &	0.2956&	0.4874 &	0.2463 \\
        & GCMC+ &	0.3968 &	0.3057 & 	0.4961 &	0.2531 \\
        & NGCF+ & 0.3814 & 0.2909 & 0.4804 & 0.2423 \\
        \cline{2-6}
        & \textbf{DGE(ours)}  &	\textbf{0.4139} &	\textbf{0.3226} &	\textbf{0.5100} &	\textbf{0.2658} \\
        \hline
    \end{tabular}    }
    \vspace{-4pt}
\end{table}

\subsection{Experimental Settings.}
\subsubsection{Experimental implementation}
For each dataset, we randomly choose 80\% data for training and the remaining 20\% data for testing following the setting in~\cite{Zhang2019}. The details about parameter settings of all the models are given in Table~\ref{t:set}. Besides the parameters varying with datasets, the settings of other parameters in baseline models are consistent with those in the original papers. We run their codes on the three datasets. We integrally train DGE with a mini-batch scheme and adopt Adam optimization approach to learn the model parameters. The hyperparameters are chosen with the minimal training loss. All the experiments are running on one machine with one TitanX GPU. For each method, we conduct the experiment for 10 times and report the average performance. 

\subsubsection{Evaluation metrics}
The relevance of predicted tags is evaluated with two typical metrics in recommendation systems including $Recall@k$ and $NDCG@k$.

\begin{figure}[tp]
    \subfigure{
    \includegraphics[width=.46\linewidth]{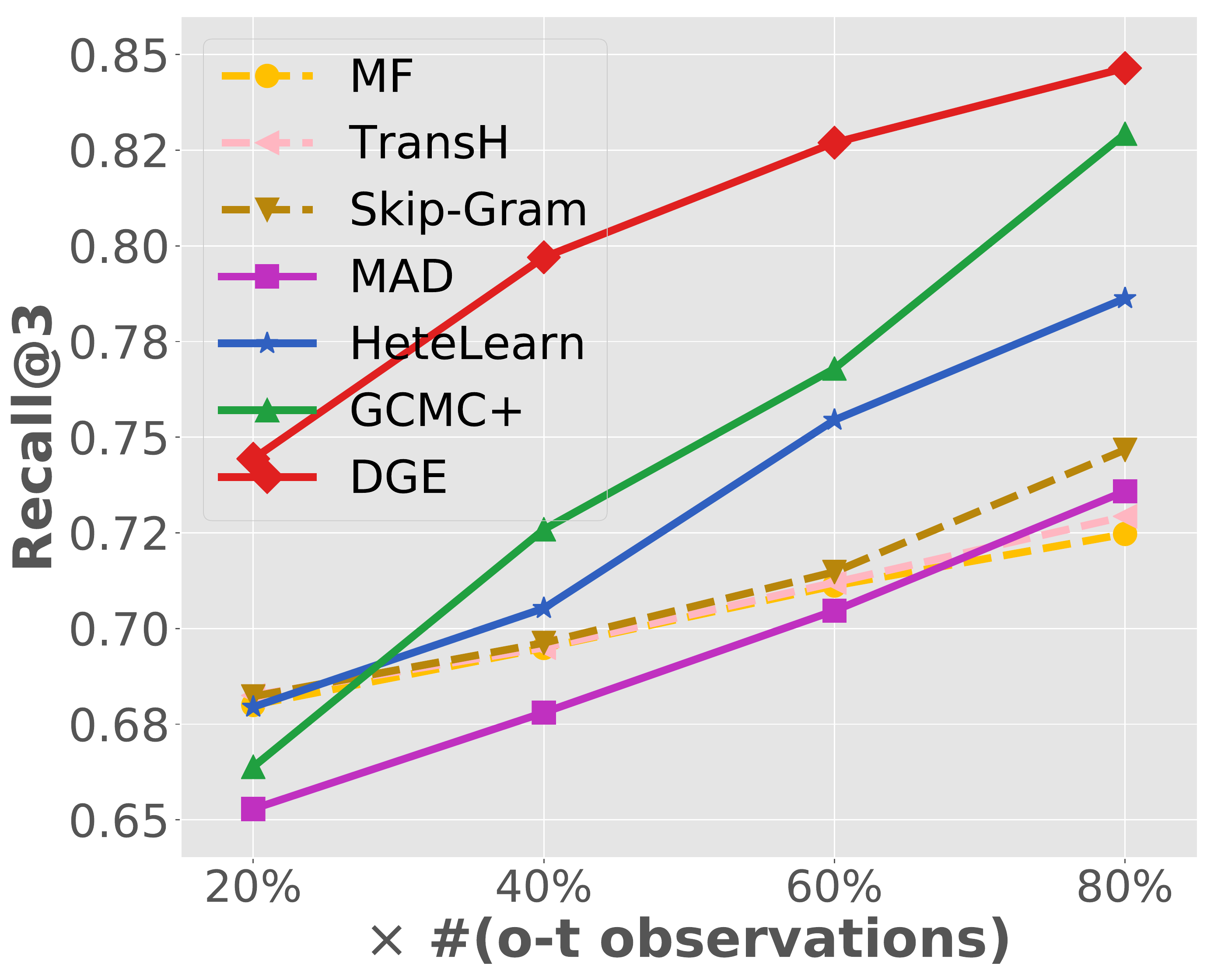}}
    \hspace{4pt}
    \subfigure{
    \includegraphics[width=.46\linewidth]{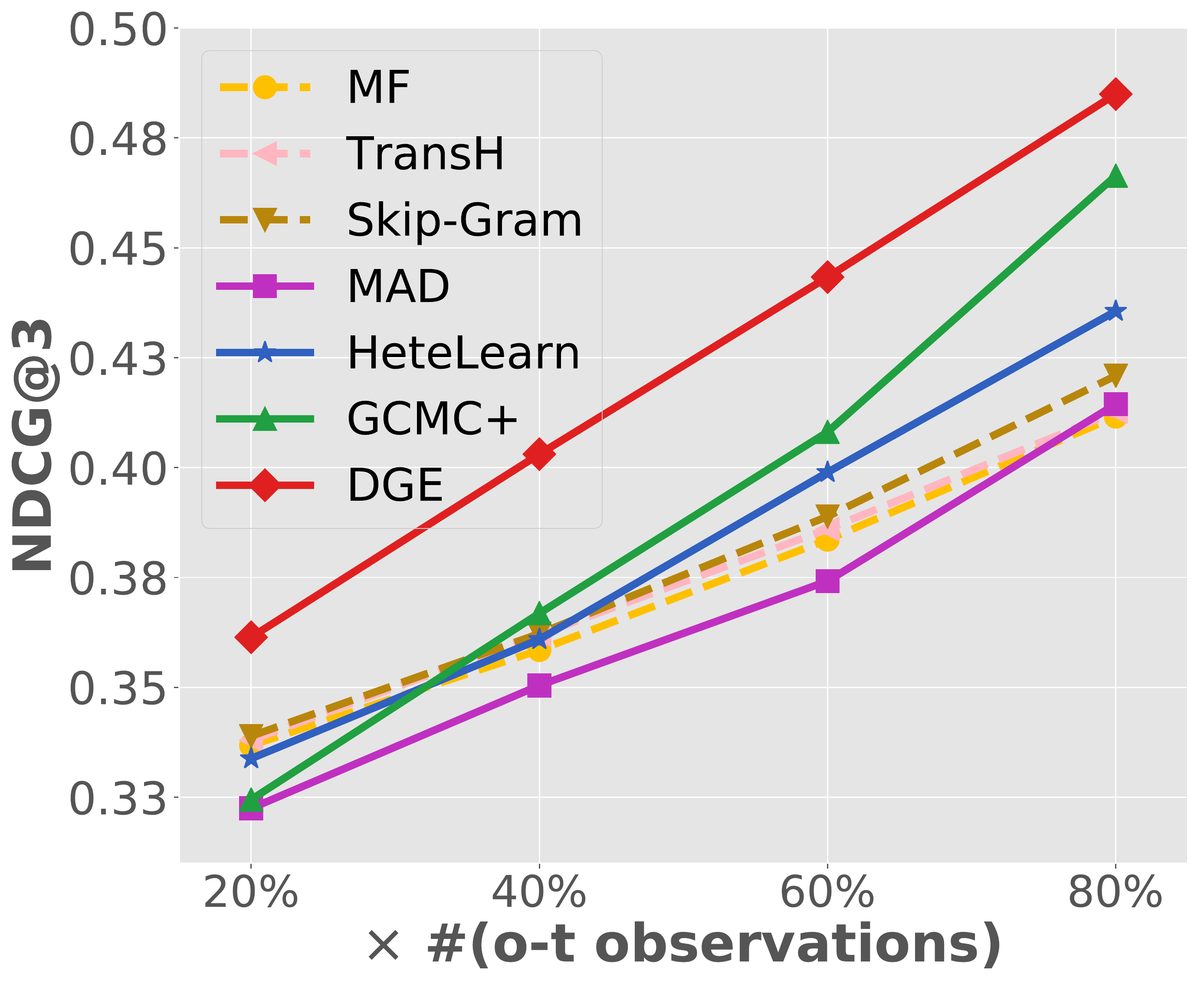}}
\centering
\caption{Recall@3 and NDCG@3 on Movie-1m with the sparsity level of object-tag observations varying. Methods denoted by dash lines only consider the first-order proximity, while those with solid lines leverage the high-order relationships in the KG.
}\label{f3}
\end{figure}

\begin{figure}[tp]
    \centering
    \includegraphics[width=.98\linewidth]{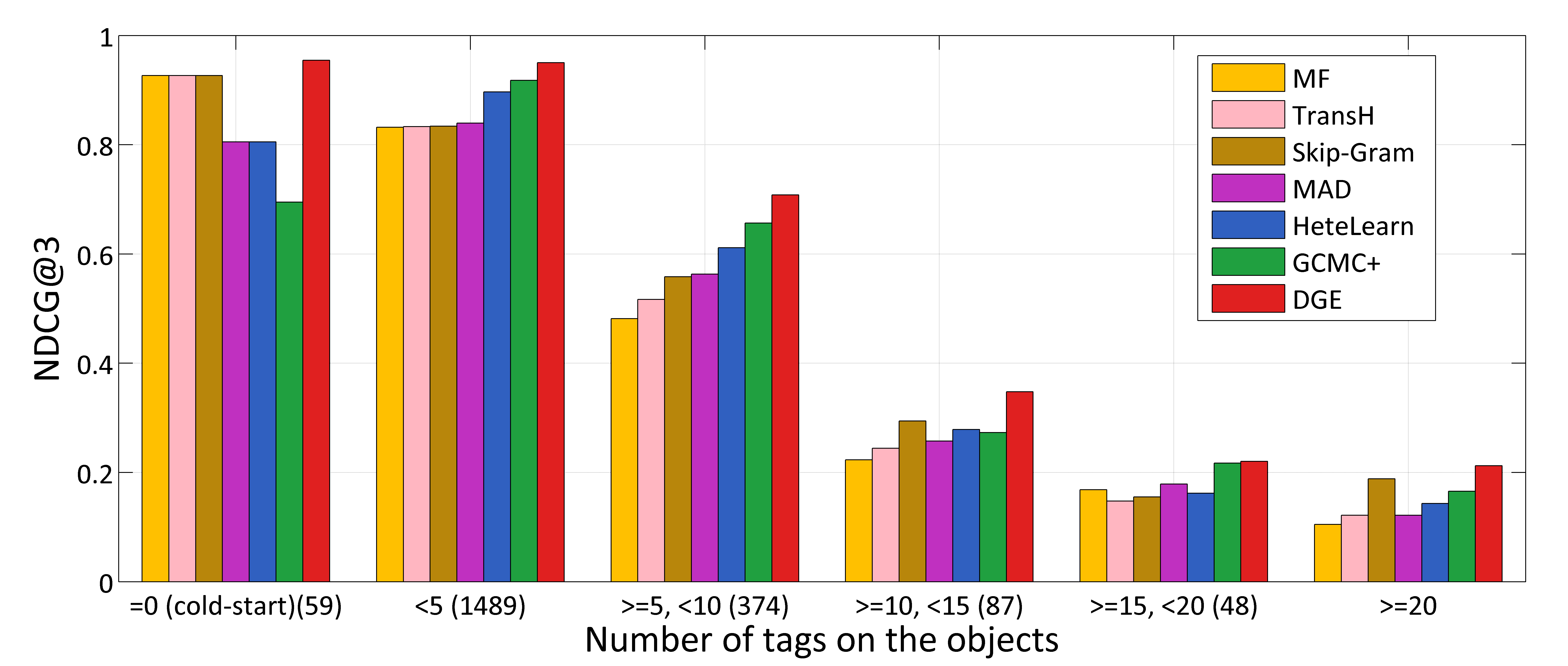}
    \caption{Recall@3 on Movielen-1m of the objects which have different number of tags, and each number in () is the number of test objects satisfying the corresponding condition.}
    \label{fig:spar_cs}
\end{figure}


\begin{table*}[t]
    \centering
    \caption{Top 5 Tags Predicted by Different Methods on Movielens-1M (The movie with * is a cold-start movie.)}
    \label{t6}
    \resizebox*{1.\linewidth}{!}{
    \begin{tabular}{|c|c|cc|cccc|}
        \hline
         \textbf{Movielens-1M} & test truth tags & TransH & Skip-Gram & MAD & HeteLearn  & \textbf{GCMC+} & \textbf{DGE} \\
        \hline\hline
         \textit{Billy Madison,} & comedy, drama & \textbf{drama}, \textbf{comedy}, & \textbf{comedy}, \textbf{drama}, & thriller, stupid,  & \textbf{drama}, \textbf{comedy},  & \textbf{drama}, \textbf{comedy}, & \textbf{comedy}, \textbf{drama},\\
         \textit{(1995)} &  & romance, adventure, & musical, romance,  & adam sandler, \textbf{drama}, &  action, romance, & war, parody, &  horror, crime,\\
          &  & action & food & \textbf{comedy} &  adventure & crime &  romance \\
        \hline
         \textit{Sense and Sensibility,} & romance, thriller & \textbf{thriller}, comedy, & \textbf{thriller}, teen movie,  & drama, based on a book, & \textbf{thriller}, comedy,  & \textbf{romance}, \textbf{thriller}, & \textbf{romance}, \textbf{thriller}, \\
         \textit{(1995)} & & adventure, action & \textbf{romance}, John Hughes, &  british, hugh grant, & \textbf{romance}, action,  & classic, witty,  & comedy, classic,\\
          & & \textbf{romance} & highschool &  romantic & adventure & fantasy & bittersweet \\
        \hline
        \textit{Madonna: Truth or Dare,} & documentary, & \textbf{drama}, \textbf{thriller}, & \textbf{thriller}, \textbf{drama},  & \textbf{thriller}, \textbf{drama}, &  \textbf{thriller}, \textbf{drama},  & \textbf{thriller}, \textbf{drama}, & \textbf{drama}, \textbf{thriller},  \\
         \textit{(1991)*} & drama, thriller & comedy, action, & action, comedy,  & comedy, action,  &  comedy, action,  & \textbf{documentary}, & \textbf{documentary},  \\
          & & romance & adventure & romance & romance  & crime, comedy &  horror, mystery \\
          \hline
    \end{tabular}
    }
    \vspace{-2pt}
\end{table*}

\begin{figure*}[!htbp]
    \subfigure[TransH]{
    \includegraphics[width=0.32\linewidth]{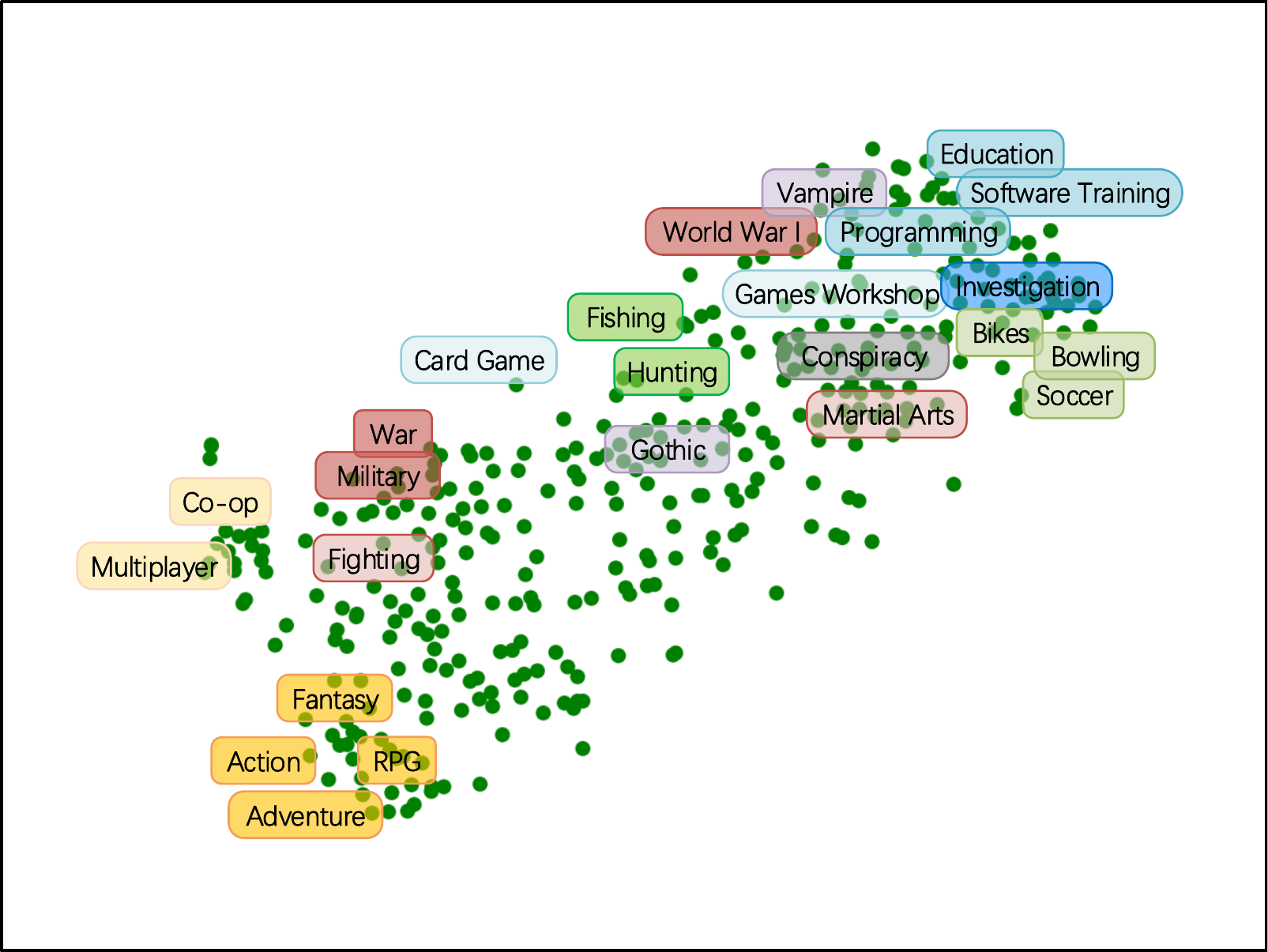}}
    \subfigure[GCMC+]{
    \includegraphics[width=.32\linewidth]{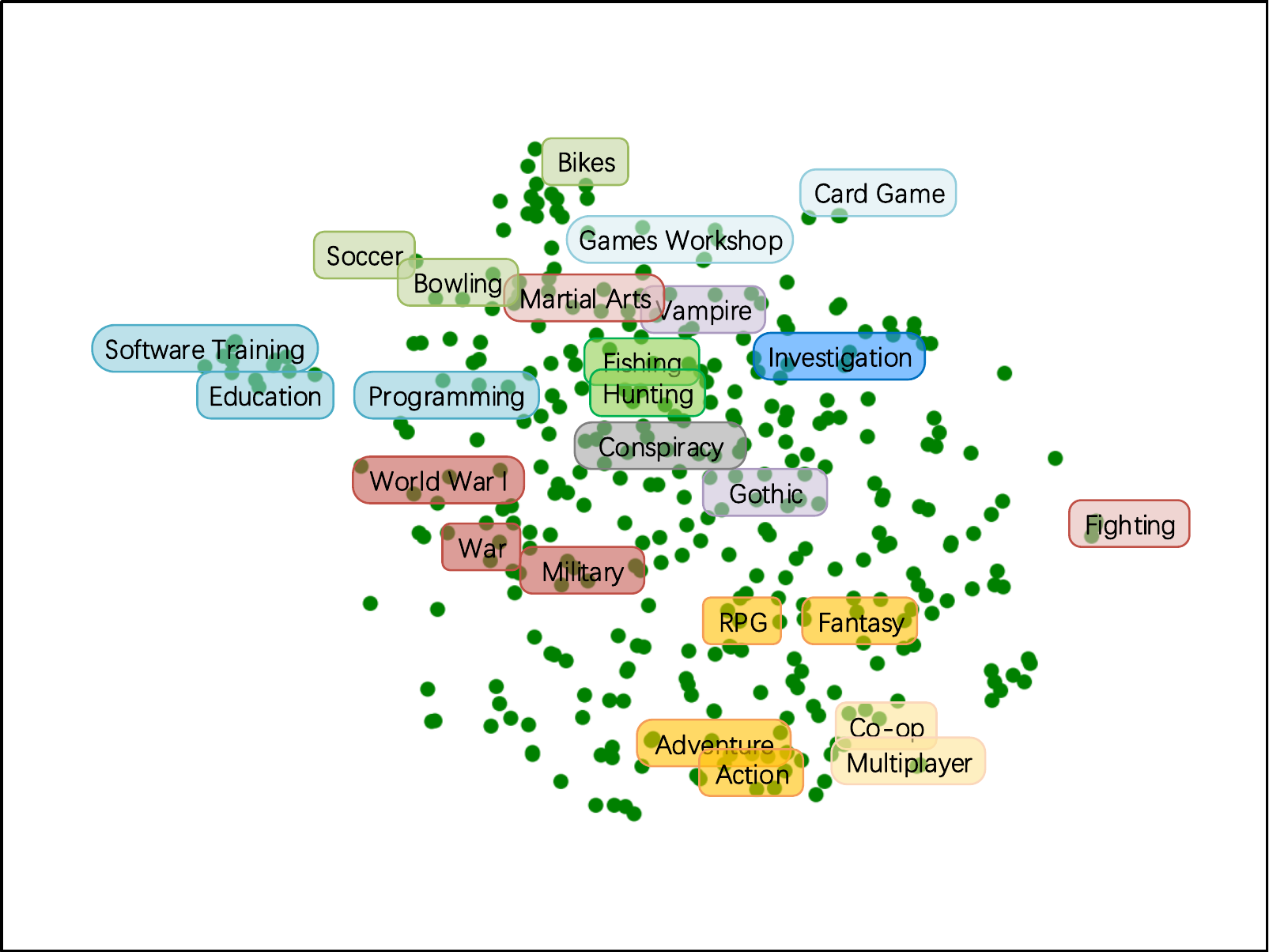}}
	\subfigure[DGE]{
		\includegraphics[width=.32\linewidth]{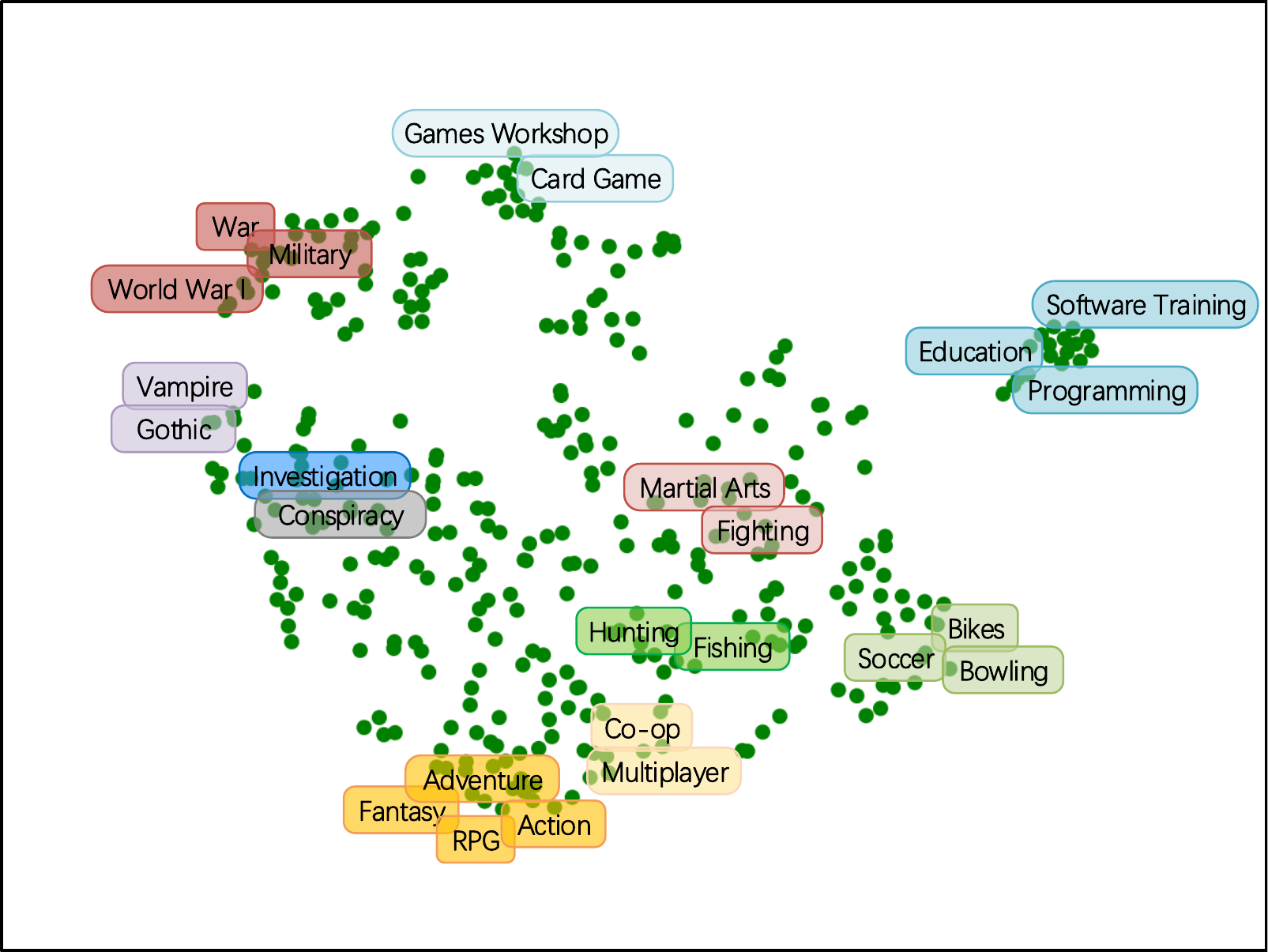}
	}
\centering
\caption{Visualization of tag embeddings derived by TransH, GCMC+ and DGE on Steam.
Tags with the same color are semantically similar. 
DGE can learn the semantic similarities between tags in the semantic space.
}\label{f4}
\vspace{-4pt}
\end{figure*}

\begin{table}[tp]
    \centering
    \caption{$Recall@k$, $NDCG@k$ on Movielens-1M, LastFM, and Steam based on Different Variants of DGE}
    \label{tab:res1}
    \resizebox*{.98\linewidth}{!}{
    \begin{tabular}{|c|c|cccc|}
        \hline
        Datasets& Methods & Recall@3 & NDCG@3 & Recall@5 &	NDCG@5 \\
        \hline \hline
        & SO-GE &0.8101	&0.4526 &	0.8401&	0.3343\\
        Movielens-1M & ST-GE &	0.7443 &	0.4207 & 	0.7823 &	0.3136 \\
        \cline{2-6}
        & \textbf{DGE}  &	\textbf{0.8464} &	\textbf{0.4850} & \textbf{0.8677} & \textbf{0.3565}\\
        \hline\hline
        & SO-GE &	0.0830&	0.0580&	0.1211&	0.0540\\
        LastFM & ST-GE &	0.1928&	0.1717&	0.2496&	0.1444\\
        \cline{2-6}
        & \textbf{DGE} &	\textbf{0.2494} &	\textbf{0.2129} &	\textbf{0.3154} &	\textbf{0.1772} \\
        \hline \hline
        & SO-GE &	0.3251&	0.2425&	0.4587&	0.2152\\
        Steam & ST-GE &	0.2856&	0.1977& 	0.4132&	0.1821\\
        \cline{2-6}
        & \textbf{DGE}  &	\textbf{0.4139} &	\textbf{0.3226} &	\textbf{0.5100} &	\textbf{0.2658} \\
        \hline
    \end{tabular}    }
\end{table}

\subsection{Overall Comparison} 
The overall performance results on the three datasets are summarized in Table~\ref{tab:res}, and we have following observations:
\begin{itemize}
\item DGE beats all other models on the three datasets, especially when predicting the top-3 most relevant tags for objects. 
The results prove that DGE extracts more essential information from both the first-order and high-order proximities in the KG to predict the missing links more accurately.
\item Comparing the methods only modeling the first-order proximity, we find that the Skip-Gram model outperforms MF and three translational distance models on all the datasets. Since the only difference between MF and Skip-Gram is the loss function, the results demonstrate that the skip-gram objective can measure the first-order proximity more accurately than the MSE loss used in MF and the margin loss used in translational distance models.
\item Compared with the methods that modeling the high-order relationship in the KG, DGE provides the most relevant tags for objects. It is because that the dual graph encoder in DGE can extract more collaborative information from the high-order proximities while the other methods are affected by some noisy information when conducting random walk or feature aggregation on the input KG.
\item DGE outperforms the Skip-Gram model. Both models adopt the skip-gram objective to learn the first-order proximity while DGE captures the high-order proximities simultaneously. The results show that the high-order proximities contain essential information for this task.
\end{itemize}


\subsection{Performance on Different Sparsity Levels} 
To evaluate the robustness of our model given the sparse object-tag observations,
we randomly draw samples (20\%, 40\%, 60\%, 80\%) of all the observed object-tag pairs for training and compare the results in Fig.~\ref{f3}. 
Experiments on all the datasets show similar results, thus we only show the results on Movielens-1M.

We find that even with a lack of tagging data, our model predicts more relevant tags than other methods. Considering that the skip-gram objective brings trivial gain in very sparse cases, DGE can still extracts supplementary information from the high-order proximities for better prediction.
Moreover, the results of MAD, HeteLearn, GCMC, and our model demonstrate that the dual graph encoder can represent both the object-object and tag-tag relations better in the sparse cases.



\subsection{Object Cold-Start and Data Sparsity Problems}
Fig.~\ref{fig:spar_cs} shows the performances on those objects which have different numbers of tags in the training set. The first group of bars corresponds to the cold-start examples. These bars show that for cold-start objects, 
DGE predicts the tags more accurately with the Recall@3 over 0.95, which verify that the high-order proximities enrich the representations of cold-start objects via the dual graph encoder. The baseline methods incorporating high-order relationships underperform the other methods in this case because that the learned models tend to accurately predict the tags for the objects in densely distributed regions (e.g. objects having $>0,<10$ tags in the training set).

Besides, by investigating the performances in the other five groups of bars in Fig.~\ref{fig:spar_cs},
we find that DGE always predicts the most relevant tags compared to other methods.
The results illustrate that DGE can mine valuable information from both the first-order and high-order proximities in the KG under different sparse cases.

\subsection{Ablation Study}
To evaluate whether the two GCN encoders in the DGE extract the high-order proximities effectively for link prediction, we replace an encoder of them with a trivial MLP. With this operation, we derive two variants of DGE :
\begin{itemize}
    
     \item SO-GE: It 
    only retains the object graph in DGE and replaces the tag GCN encoder with MLP. 
     This model cannot extract the high-order proximity between tags.
    
    \item 
    ST-GE: It 
    only retains the tag graph in DGE and a MLP for objects is applied to derive object embeddings containing no the high-order proximity information.
\end{itemize}
We compare the results on three datasets via these variants and DGE in Table.~\ref{tab:res1}. The results 
prove that the designed dual graph encoder can learn the helpful structural information in the high-order object-object and tag-tag proximities to enhance the prediction performances.






\subsection{Visualization}
We first apply TSNE to the high-dimension tag embeddings derived by TransH, GCMC+ and DGE. The visualization results are shown in Fig.~\ref{f4}. 
Compared to TransH and GCMC+, tag embeddings derived by DGE can represent the semantic similarities between tags more clearly, which improves the interpretability of the prediction results. For example, the semantically similar
tags \textit{``Vampire"} and \textit{``Gothic"} are close in Fig.~\ref{f4}~(c) but far apart in  Fig.~\ref{f4}~(a) and (b).
Besides, DGE clusters tags into multiple classes in the semantic space more explicitly than the other two methods. The results prove that our model can learn the semantic similarities between tags via embedding the high-order proximity between tags explicitly.
Accordingly, our model will predict tags that are more probable to be semantically relevant to the target object.

In addition, we give three tagging examples on Movielens-1M in Table~\ref{t6}. 
We find that for the cold-start movie \textit{``Madonna: Truth or Dare, (1991)"}, the former four methods provide two most popular tags \textit{``drama"} and \textit{``thriller"} without representing the object-object proximity sufficiently, while DGE predicts the accurate tags. This result shows that our model can alleviate the object cold-start problem. 
For the movie \textit{``Sense and Sensibility, (1995)''}, the Skip-Gram model predicts the tags more accurately than TransH, proving that the skip-gram objective can better learn the first-order object-tag proximity.
Besides, for the latter two movies,
our model puts the most relevant tags at the top of the lists, which further proves the prediction accuracy of DGE.

\section{Conclusions} \label{sec:6}
In this paper, we propose a Dual Graph Embedding (DGE) method in an auto-encoding architecture 
to capture the first-order and high-order proximities in the input KG
for the object-tag link prediction task. Here the dual graphs include the object and tag graphs that are built to depict the high-order proximities.
Then the encoder 
embeds the two types of high-order proximities in the dual graphs into object and tag embeddings. The decoder models the first-order proximity between objects and tags over the global proximity structure from the skip-gram perspective. Under the supervision of the decoder, the similarity information from both the first-order and high-order proximities is extracted for better prediction.

\bibliographystyle{IEEEtran}
\bibliography{mybib2}

\end{document}